\documentclass{article}
\usepackage{a4,graphicx,epsfig}
\def\question#1 {~\\ {\bf\it #1 }\\}
\def\question#1 {}
\newcommand{\A}{{\mathcal A}}
\newcommand{\p}{{\mathcal A}}
\newcommand{\be}{\begin{equation}}
\newcommand{\ee}{\end{equation}}
\newcommand{\ba}{\begin{eqnarray}}
\newcommand{\ea}{\end{eqnarray}}
\newcommand{\bb}{\begin{thebibliography}}
\newcommand{\eb}{\end{thebibliography}}

\newcommand{\ua}{\sigma^{(\uparrow)}}

\newcommand{\da}{\sigma^{(\downarrow)}    }

\newcommand{\ahnf}{${\mathcal A}^h_{nf}$}
\newcommand{\acsf}{${\mathcal A}^{em}_{sf}$}

\begin{document}

\begin{center}{ \large\bf
Hadron spin-flip amplitude: \\an analysis of the new
$A_N$ data from RHIC\footnote{To appear in the proceedings of
the conference ``Spin and Symmetry'', Prague,
              July 12-19 (2003). }} \\
\bigskip
{\underline{O.V. Selyugin}\footnote{on leave from Bogoliubov
 Laboratory of Theoretical Physics, JINR, 141980, Dubna, Moscow Region,
 Russia. \\
emails: selugin@qcd.theo.phys.ulg.ac.be, \ J.R.Cudell@ulg.ac.be
} 
 and J.-R. Cudell}\\ \medskip

Institut de Physique, B\^at. B5a, Universit\'e de Li\`ege\\
 Sart Tilman, B4000 Li\`ege, Belgium 
\end{center} 
\bigskip
\centerline{\bf Abstract}
Through a direct analysis of the scattering amplitude,
we show that the preliminary measurement of $A_N$ obtained by the E950 
Collaboration at different energies are mainly sensitive
to the spin-flip part of the amplitude in which
the proton scatters with the $^{12}C$ nucleus as a whole. 
The imaginary part of this amplitude is negative,
and the real part is positive 
and has a large slope. 
We give predictions for 
$p_L = 600$~GeV/$c$, which  depend mainly on
the size 
and energy dependence
of the real part of the amplitude. \\ \bigskip

The new RHIC fixed-target data, from E950, consists 
in measurements of the analysing power 
\be A_N(t)={\ua-\da\over \ua+\da}\ee
for momentum transfer $0 \leq |t| \leq 0.05$~GeV$^2$,
for a polarised $p$ beam hitting a (spin-0) $^{12}C$. In this 
region of $t$, 
the electromagnetic amplitude is of the same order of magnitude 
as the hadronic amplitude, and the interference of the imaginary part 
of \ahnf $ \ $ with the spin-flip part of the 
electromagnetic amplitude \acsf $ \ $ leads to a peak in the analysing 
power $A_N$,
usually referred to as the Coulomb-Nuclear Interference (CNI) effect
\cite{schwinger}.

The first RHIC measurements at $p_L = 22 $~GeV/$c$ \cite{an22}  
in $p ^{12}C$ scattering indicated however that
$A_N$ may change sign already at very small momentum transfer.
Such a behaviour cannot be described by the CNI effect alone. Indeed,
 it also requires some contribution of the hadron spin-flip amplitude.
  The first analysis of the preliminary data \cite{kt} gives for the ratios of
  the real and imaginary parts of the reduce spin-flip amplitude
  to the imaginary part of the spin non-flip amplitude 
  $    R = 0.088 \pm 0.058$ and $I = -0.161 \pm 0.226 $, with the reduced
spin-flip amplitude being defined as
$ \tilde{\A}_{sf}(s,t) \equiv  2\ m_p\ \A_{sf}(s,t)/ \sqrt{|t|}$.
The large error on $I$ unfortunately 
leads to a high uncertainty on the size of 
the hadronic spin-flip amplitude. 

Isoscalar targets such as $^{12}C$ simplify the calculation as they 
suppress the contribution of the isovector reggeons $\rho$ and $a_2$ 
by some power of the atomic number. Also, as $^{12}C$ is spin 0,
there are only two independent helicity amplitudes: proton spin flip 
and proton spin non flip.
However, nuclear targets lead to large theoretical uncertainties
because of the difficulties linked to nuclear structure, and 
because of the lack of high-energy proton-nucleus scattering 
experiments.
Given these problems, we shall not rely on theoretical models
(such as the Glauber formalism) but rather parametrise the scattering 
amplitude directly from data, and take 
the interference terms fully  into account in the form of the analysing power.

The elastic and total cross sections and the analysing power 
$A_N$ for $p ^{12}C$ scattering are given by
\begin{eqnarray}
d\sigma/dt \ &=& \   2 \pi \left(|\A_{nf}|^2+ |\A_{sf}|^2\right), 
 \nonumber\\
\sigma_{tot}&=& 4 \pi Im(\A_{nf}),\label{an}\\
A_N \ d\sigma/dt \ &=&  \ - 2\pi 
                   Im[ \A_{nf} \A_{sf}^{*})]. \nonumber
\end{eqnarray}
 
Each term includes a hadronic and an electromagnetic contribution:
$
 \A_i(s,t) = \p^h_{i}(s,t) 
        + \p_{i}^{em}(t) e^{i\delta}, (i=nf, sf),
$
where $\p^h_{i}(s,t)$ describes the strong interaction of $p ^{12}C$,
and $\p_{i}^{em}(t)$  the  electromagnetic interaction.
$\alpha_{em}$ is the electromagnetic fine structure constant, and
the Coulomb-hadron phase $\delta$ is given by
$\delta=Z\alpha_{em} \varphi_{CN}$ 
with $Z$ the charge of the nucleus, and $\varphi_{CN}$
the Coulomb-nuclear phase \cite{prd-sum}.
The electromagnetic part of the scattering amplitude can be written 
as
\begin{eqnarray}
\A^{em}_{nf}&=& {2 \alpha_{em} \ Z\over t} \ F^{^{12}C}_{em} F^{p}_{em1},
 \\ \nonumber
\A^{em}_{sf}&=& - {\alpha_{em} \ Z\over m_p \sqrt{|t|}} \ 
                           F^{^{12}C}_{em} F^{p}_{em2},
\end{eqnarray}
where 
$ F^{p}_{em1}$ and $ F^{p}_{em2}$ are
the electromagnetic form factors of the proton, and
$F^{^{12}C}_{em}$ that of $^{12}C$.
We obtain $F^{^{12}C}_{em}$
from the electromagnetic density of the nucleus. 

\begin{figure}
\epsfysize=8.cm
\epsfxsize=12.cm
\vglue -1.cm
\centerline{\epsfbox{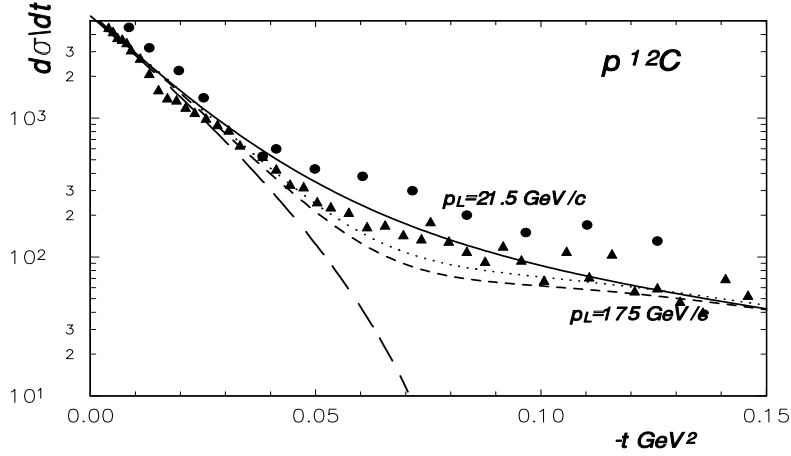}}
\caption{
$d \sigma /dt $ calculated by Glauber model 
  \cite{kt}(formula (37),
  the previous plus the maximal possible inelastic $d\sigma/dt$, 
 which fall with   $t \rightarrow 0$,  first plus inelastic 
 without the decreasing coefficient, and by our formula(9) 
 (long-dashed, dashed, doted, solid curves
  correspondingly). The experimental data: circles - \cite{bell}, triangles -
  \cite{shiz}.   }
 \end{figure}

The parts of the scattering amplitude due to strong interaction 
are assumed to be  approximated by falling exponentials in the 
small-$t$ region. 
The slope parameter $B(s,t)/2$ is then the derivative 
of the logarithm of the amplitude with respect to $t$.  
If one considers only one contribution to the amplitude, 
this coincides with 
the slope of the differential cross section. In a more
complicated case,
there is no direct correspondence with the
cross section because of interference terms.   

  The quick growth of the differential cross sections in elastic
  hadron-nucleus scattering, which is reflected by the large slope at
  small momentum transfer, does not permit to make  a hard  theory input.
  Pure Glauber theory gives in this region a behaviour for the slope
  which is very close to that obtained in the black disk limit: 
  the slope increases with $|t|$, in accordance with
  low-energy data. However, at high energies,
   the slope slightly increases as $t\rightarrow 0$.
For example, in the data of \cite{shiz}, the slope of
   the proton-Carbon elastic cross section slightly increases when 
   $|t| \rightarrow 0$ and at $|t| \approx 0.01 \ $GeV$^2$ is equal to
   $B= 74 \ $GeV$^{-2}$ at $p_L=170 \ $GeV$/c$. More recent data \cite{selex}
   give $B= 60\ $GeV$^{-2}$ at $p_L= 600 \ $GeV$/c$ in the region 
   $0.01 \leq |t| \leq 0.03 \ $GeV$^2$.

     The description of part of the experimental data by different
   approaches is shown in Fig.1.   The Glauber model gives a sharp
   minimum at $|t| \approx 0.1 \ $GeV$^2$. An additional contribution to
   the differential cross section, coming from quasi-elastic scattering,
   cannot remove the effect of this sharp minimum in this region.
   The data of \cite{selex}, as the old data \cite{shiz}, show a smooth change
   of the slope at small momentum transfer. 
     Hence, if we want to obtain a good description of the diffraction
   peak in proton-Carbon scattering, we need to take a more complicated
   form of the scattering amplitude than given by the Glauber model 
   or the simple exponential behaviour. 
    Thus, for our  phenomenological analysis
   of the preliminary data,  we use the representation of the hadron
   spin non-flip amplitude in the form of two exponentials \cite{d1,prc51}.
   It gives us a large slope at very small momentum transfer
   which smoothly changes at higher $|t|$.
   We shall not  discuss here the origin of these exponentials. 
   Note that they can be connected with different 
  interaction mechanisms at small and large distances.
  
  \begin{eqnarray}
{\mathcal A}^{h}_{nf}(s,t) \ &=& \ \A^{h1}_{nf}(s,t) \ 
+ \   \A^{h2}_{nf}(s,t),\label{full},\\
\A^{h1}_{nf}(s,t) &=& (1+\rho^{h1}) 
{\sigma_{tot}^{h1}(s)\over 4\pi} \exp\left({B^{h1}\over 2}t\right), \\
\A^{h2}_{nf}(s,t) &=& (1+\rho^{h2})  
{\sigma_{tot}^{h2}(s)\over 4\pi} \exp\left({B^{h2}\over 2}t\right). 
\label{f-h}
\end{eqnarray}
  where $B^{h1} > B^{h2}$ and the amplitude $\A^{h1}_{nf}(s,t)$ gives
  the main contribution to the differential cross sections when
  $t \rightarrow 0$.
  For simplicity, 
  we take the additional amplitude  $\A^{h2}_{nf}(s,t)$
   with small slope as proportional to the
  nucleon-nucleon total cross section. The normalisation is determined by
  comparison of our calculation with the experimental data
  at $-t=0.1 \ $GeV$^2$,
  where the Glauber model has the first minimum in the differential 
 cross sections.    We neglect
  the possible contribution of inelastic effects, as 
  we do not know its size and as at small $t$ it must decrease,
  see for example \cite{GM}. Our $d \sigma/dt$ is a little lower
than the data of 
  \cite{bell}, indicating where possible inelastic contributions might set in. 
  The coincidence of the hard and electromagnetic form factors 
  at small $t$ supports our representation. 
   
The experiment \cite{selex} on $pC$ scattering at $p_L = 600$~GeV/$c$ 
gives us 
  $\sigma^{pC}_{tot} \ = \ 341 $~mb;  $B^{pC}(t \approx 0.02{\ \rm GeV}^2)
 =  62$~GeV$^{-2}$.
  To obtain the values of these parameters for other energies, 
we make the following 
assumptions on their energy dependence:
some analysis \cite{karol} and the data \cite{murthy}
show that the ratio $R_{C/p}$ of $\sigma_{tot}(p ^{12}C)$ to $\sigma_{tot}(pp)$
decreases very slowly in the region $5 \leq p_L \leq  600 $~GeV/$c$.
We take its energy
dependence, according to the data~\cite{shiz}, as 
$ R_{C/p} = 9.5 \ (1 - 0.015 \ln{s})$.
From this we obtain $\sigma_{tot}^{h1}(s)\equiv\sigma_{tot}^{pC}(s) 
- \sigma_{tot}^{h2}(s)
$.
We assume that the slope slowly rises with $\ln{s}$ in a way 
similar to the $pp$ case, and normalise 
 it so that the full amplitude (\ref{full}) has a slope of  
 $62$~GeV$^{-2}$ at $p_L= 600$~GeV/c and
 $|t|=0.02$~GeV$^2$. This gives $ B^{h1}= 70 \ (1 +0.05 \ \ln{s})$.

We do not know the energy dependence of $\rho^{h1}$, 
but because the $\rho$ and $a_2$ trajectories are suppressed, and
because they contribute negatively, it
must be higher than in the $pp$ case, where it is about $-0.1$ in 
this energy region.
In fact, 
the data from \cite{selex,shiz} indicate that $\rho^{h1}$ can be positive. 
We also
know that at very high energy, $\rho^{h1}$ should be of the order of
$\rho^{pp}$, which is about 0.1. 
We thus assume that, at RHIC energies, it is of the order of 0.05, 
and that it changes
logarithmically with $s$, similarly to the $pp$ case.
We also allow for an extra term proportional 
to $\rho_{pp}$
with a linear suppression in $A$. This gives us two variants:  
\begin{eqnarray}
\rho^{h1}=\rho_{I}^{h1}& 
   =& 0.05/(1- 0.05 \ \ln{s}) + \rho_{pp}/A, \label{fro} \\ 
\rho^{h1}=\rho_{II}^{h1}& =& 0.05/(1- 0.05 \ \ln{s}). \label{sro}
\end{eqnarray}   

We parametrise the 
spin-flip part of $p ^{12}C$ scattering also by two exponents    
\begin{eqnarray}
{\A^{h}_{sf}}(s,t)&=&   (k_2\rho^{h1}+i k_1) 
  { \sqrt{|t|}\sigma_{tot}^{h1}(s)\over 4\pi} 
\exp\left({ B^{h1}\over 2}t\right)\nonumber, \\
 {\tilde\A^{h2}_{sf}}(s,t)&=&{{\A^{h2}_{nf}}(s,t)/ 10} .\label{sfpN}  
\end{eqnarray}     
We have assumed here that 
the spin-flip and the spin-non-flip amplitude have the same slope.
One could of course allow for more freedom and take different slopes,
but the data are not yet precise enough to test this.

\begin{figure}
\epsfysize=6.cm
\epsfxsize=8.5cm
\vglue -1cm
\centerline{\epsfbox{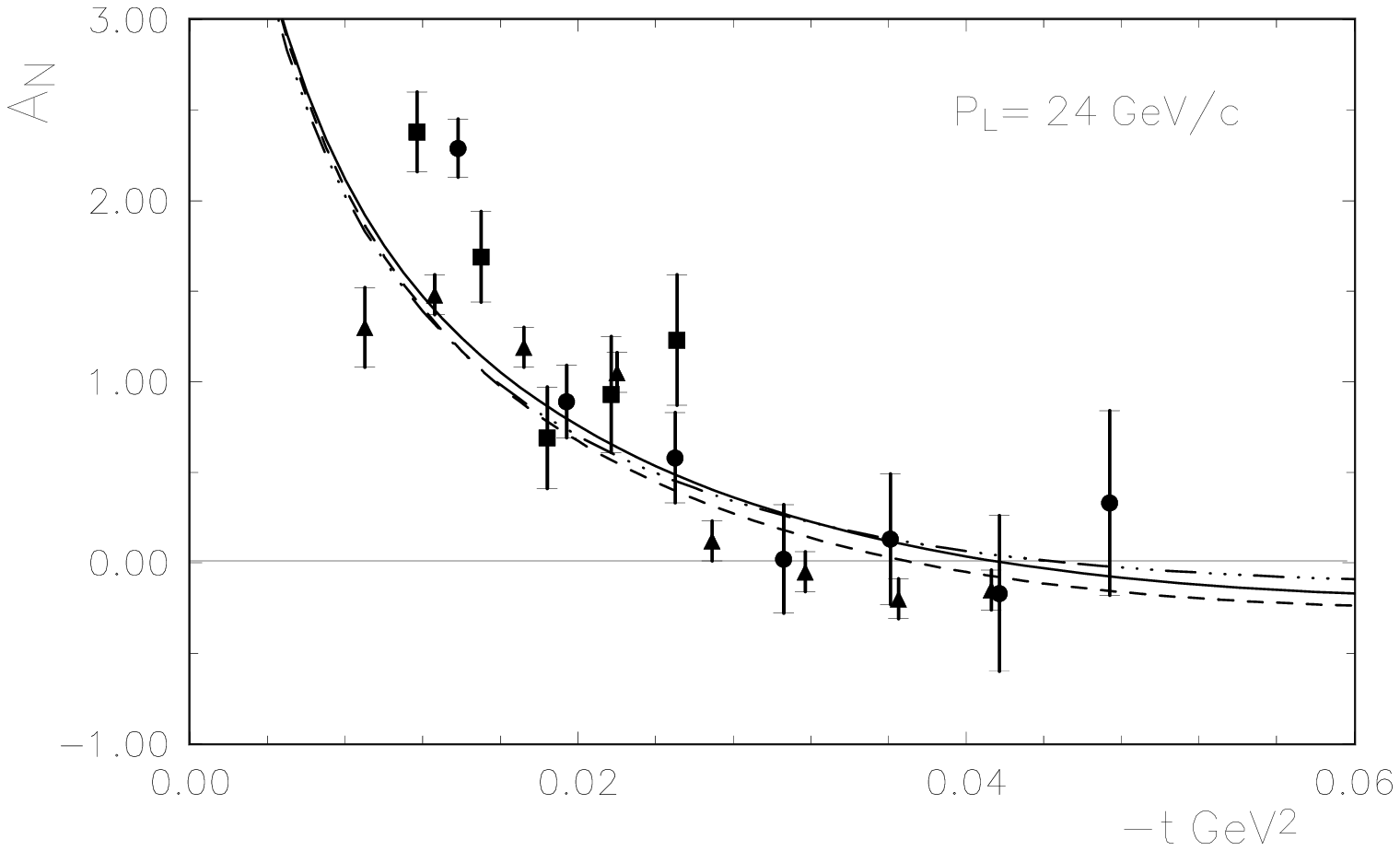}}
\vglue -1cm
\centerline{\epsfbox{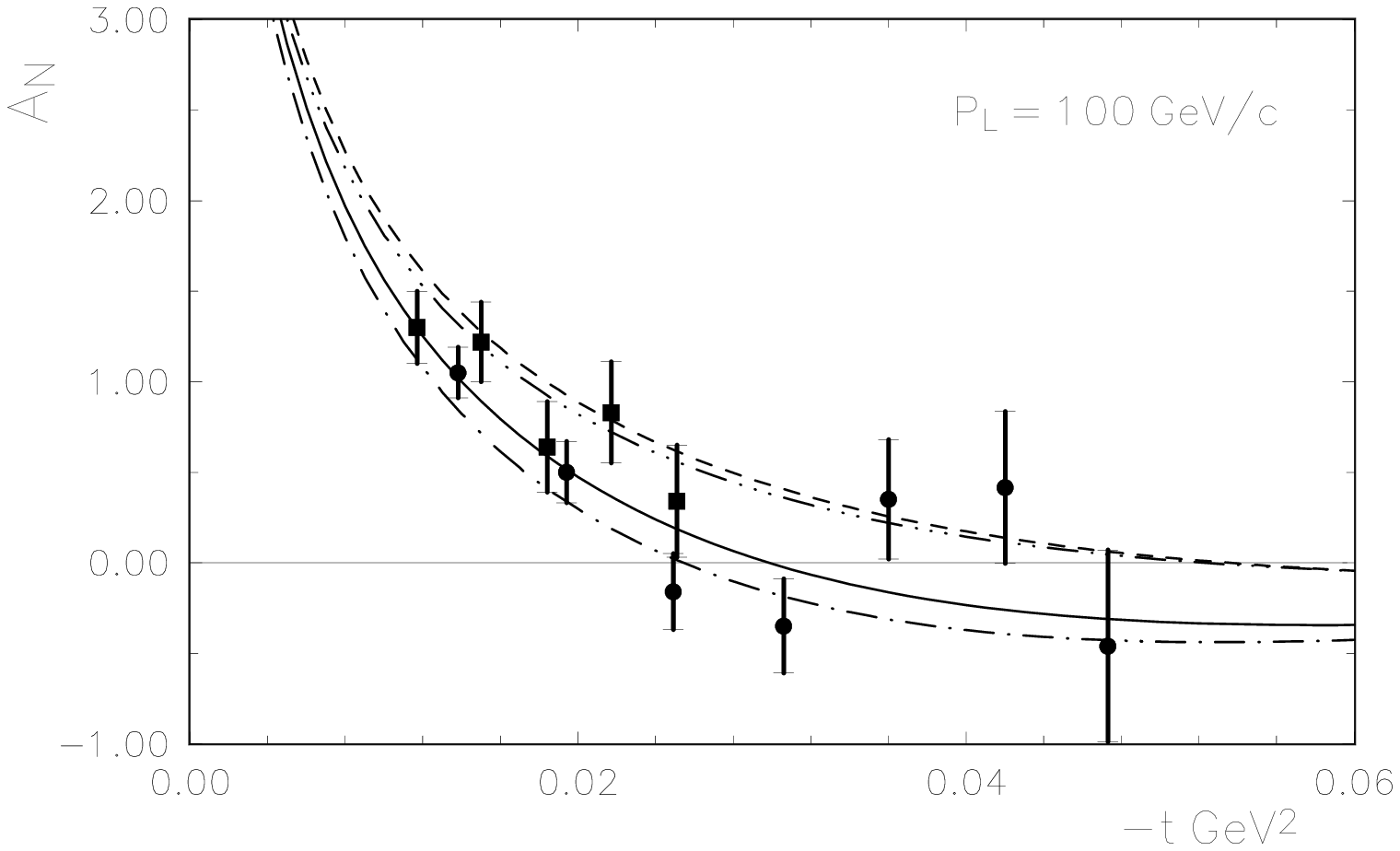}}
\caption{ (a) and (b) $ \ \ \ \ $
The analysing power    $A_N $ (in \%) at $p_{L} = 24$~GeV/$c$ (a) and
$100 $~GeV/$c$ (b), compared with the data 
\cite{an22,an100} (only statistical errors are shown). 
The two scenarios (\ref{fro}) and (\ref{sro}) lead respectively to
the upper and lower curves (these are indistinguishable at 24~GeV/$c$).
 The dot-dashed curves correspond to the addition of a small
 spin-flip contribution from  (\ref{sfpN}).
}
\end{figure}

From the full scattering amplitude, the analysing power is given by  
\begin{eqnarray}
  A_N\frac{d\sigma}{dt} =
         - 4 \pi [Im(\A_{nf})Re(\A_{sf})-Re(\A_{nf})Im(\A_{sf})],\nonumber 
\end{eqnarray} 
each term having electromagnetic and hadronic contributions.
We can now calculate the form of the analysing power $A_N$ at
small momentum transfer with different 
coefficients $k_1$ and $k_2$ chosen to obtain the best description
of $A_N$ at $p_L=24 $~GeV/$c$ and $p_L = 100 $~GeV/$c$. 
Of course, we only aim at a qualitative description
as the data are only preliminary and as they are
normalised to those
at $p_L = 22 $~GeV/$c$ \cite{an22}.

The preliminary data show that $A_N$ decreases
very fast after its maximum and is almost zero in a large
region of momentum transfer. This behaviour can be explained
only if one assumes a negative contribution of the interference 
between different parts of the hadron amplitude, that    
changes slowly with energy. 
\begin{figure}
\epsfysize=6.cm
\epsfxsize=9.cm
\vglue -1.cm
\centerline{\epsfbox{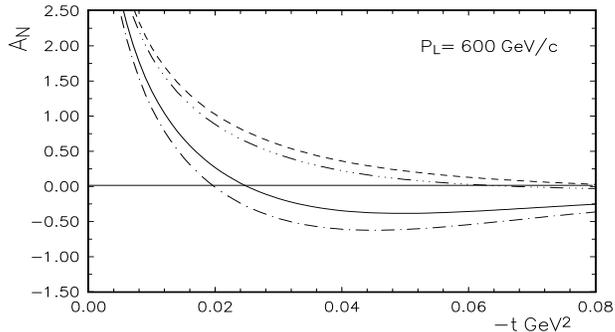}}
\caption{
  The  predictions 
for   $A_N $ (in \%) at $p_{L} = 600 $~GeV/$c$. The curves are as in Fig.~2.}
\end{figure}

 The data at  $p_L = 100 $~GeV/$c$ decrease faster
than those at $p_L = 24 $~GeV/$c$, and the zero 
of $A_N$ moves to lower values of $|t|$.
This change of sign is independent from the normalisation 
of the data. It would be very interesting to
obtain new data with higher accuracy and at higher energies in order to
distinguish between the two scenarios (\ref{fro}) and (\ref{sro})
 (see Fig. 3).
Both variants give the same size and the same negative sign 
for the imaginary part 
of the spin-flip amplitude. 
As mentioned above, such an amplitude gives
an additional positive contribution to the CNI-effect at the maximum.
Its size is mostly determined by the magnitude of $A_N$ at small $|t|$.
The fast change of sign of $A_N$ is explained by the 
interference of different parts of the hadronic amplitude.

Hence, the shape and energy dependence of the analysing power 
depend mostly on the size and energy dependence of 
$\rho_{pC}$. If we choose  another size and energy dependence, 
we can obtain a different shape for $A_N$ and different 
magnitudes for $k_1$ and $k_2$. However all conclusions
will stand and  $I=Im(r_5)$ will remain negative. Note that a positive
$I$ with our choice of $\rho$
  would lead to an increase with energy of the value of $p_L$
at which $A_N$ has a zero.
    
The ratio of the reduced spin-flip 
amplitude to the spin-non-flip amplitude is approximately 
$15 \%$. 
Note that the description of the experimental data and 
of their energy dependence
is heavily correlated with the size of the slope of the spin-flip amplitude.
We obtain a slope equal to $85.5 \ $ GeV$^{-2}$ for 
 $p_L = 24 \ $GeV/c and  $94.6 \ $ GeV$^{-2}$ for $p_L = 600 \ $GeV/c.

~\bigskip 

\noindent{\it Acknowledgements:} O.V.S. is a Visiting Fellow of the
Fonds National pour la 
Recherche Scientifique, Belgium. We thank V.~Kanavets and D.~Svirida
for their comments and discussions.

\end{document}